\documentclass[a4paper, 12pt]{article}
\usepackage{epsfig}
\usepackage{graphicx}
\usepackage{amsmath}

\author{
Juli\'an Candia$^{a,b}$, Marta C. Gonz\'alez$^{a,b}$, Pu Wang$^{a,b}$,\\
Timothy Schoenharl$^{c}$, Greg Madey$^{c}$, Albert-L\'aszl{\'o} Barab\'asi$^{a,b,d}$\\{}\\
$^a${\small\it Center for Complex Network Research and Department of Physics,}\\  
{\small\it Northeastern University, Boston, MA 02115, USA}\\
$^b${\small\it Department of Physics, University of Notre Dame, Notre Dame, IN 46556, USA}\\
$^c${\small\it Department of Computer Science and Engineering,}\\
{\small\it University of Notre Dame, Notre Dame, IN 46556, USA}\\
$^d${\small\it Collegium Budapest, Szenth\'aroms\'ag u. 2, H-1014 Budapest, Hungary} 
}  

\title{Uncovering individual and collective human dynamics from mobile phone records}

\begin{document}
\maketitle

\begin{abstract}
Novel aspects of human dynamics and social interactions are investigated by means of 
mobile phone data. Using extensive phone records resolved in both time and space, we 
study the mean collective behavior at large scales and focus on the occurrence of anomalous events. 
We discuss how these spatiotemporal anomalies can be described using standard percolation theory tools.
We also investigate patterns of calling activity at the individual level and show 
that the interevent time of consecutive calls is heavy-tailed. This finding, which has 
implications for dynamics of spreading phenomena in social networks, agrees with results previously 
reported on other human activities.  
\end{abstract}

\section{Introduction}
Mobile phones are becoming increasingly ubiquitous throughout large 
portions of the world, especially in highly populated urban areas and 
particularly in industrialized countries, where mobile phone penetration 
is almost $100\%$. 
Mobile phone providers regularly collect extensive data about the call volume, calling 
patterns, and the location of the cellular phones of their subscribers. In order for 
a mobile phone to place outgoing calls and to receive incoming calls, it must 
periodically report its presence to nearby cell towers, thus registering its position 
in the geographical cell covered by one of the towers.   
Hence, very detailed information on the spatiotemporal 
localization of millions of users is contained in the extensive call records of any 
mobile phone carrier. If misused, these records - as well as similar datasets on 
buying habits, e-mail usage, and web-browsing, for instance - certainly pose a serious 
threat to the privacy of the users. However, the use 
of privacy-safe, anonymized datasets represent a huge scientific opportunity to uncover the structure and 
dynamics of the social network at different levels, from the small-scale individual's 
perspective to the large-scale, collective behavior of the masses, with an unprecedented 
degree of reach and accuracy. 
Besides the inherent scientific interest of these issues, 
deeper insight into applications of great practical importance could certainly be gained. For instance, 
urban planning, public transport design, traffic engineering, disease outbreak control, and 
disaster management, are some areas that will greatly benefit from a better understanding 
of the structure and dynamics of social networks \cite{gon07}.    

The use of mobile phone data as a proxy for social interaction has already proved 
successful in several recent investigations. Onnela {\it et al.} \cite{onn07a,onn07b} have 
analyzed the structure of weighted call graphs arising from reciprocal calls 
that serve as signatures of work-, family-, leisure- or service-based relationships. A coupling between 
interaction strengths and the network's local structure was observed, with the counterintuitive consequence 
that social networks turn out to be robust to the removal of the strong ties but fall apart following a 
phase transition if the weak ties are removed. Szab\'o and Barab\'asi \cite{sza06} have studied social network 
effects in the spread of innovations, products and new services. They investigated different mobile 
phone-based services and found the coexistence on the same social network of two distinct usage classes, 
with either very strong or very weak community-based segregation effects. 
In the context of urban studies and planning, Ratti {\it et al.} \cite{ratti1,ratti2} have considered  
the potential use of aggregated data from mobile phones and other hand-held
devices. Their ``Mobile Landscapes" 
project aims at the application of location based services to urban studies in order to gain insight into 
complex and rapidly changing urban dynamics phenomena.       
More recently, Palla, Barab\'asi and 
Vicsek \cite{pal07a, pal07b} used mobile phone data to study the evolution of social groups. They found that 
large groups persist for longer times if they are capable of dynamically altering their membership, suggesting 
that an ability to change the group composition results in better adaptability. In contrast, the behavior of small 
groups displays the opposite tendency, the condition for long-term persistence being that their composition remains stable.      

In the following sections, we present new results that 
address novel aspects of human dynamics and social interactions obtained from extensive mobile phone data. 
In Sect. 2 we show how large-scale collective behavior can be described using aggregated data 
resolved in both time and space. We stress the importance of investigating large departures from the average and 
develop the basic framework to quantify anomalous fluctuations by means of standard percolation theory tools.   
In Sect. 3 we focus on the individual level and study patterns of
calling activity. We show that the interevent time of consecutive
calls is heavy-tailed, a finding that has implications for the
dynamics of spreading on social networks~\cite{pas01,Zoltan,Grenfell:Science,Vespignani,gon_epi1,gon_epi2,
can06,can07a,can07b}. Furthermore, by 
fixing the time of observation between consecutive 
calls it is possible to use the phone call data
to characterize some aspects of human mobility.  

\section{Fluctuations in aggregated spatiotemporal call activity patterns}

The spatial dependence of the call activity at any given time can be conveniently displayed by 
means of maps divided in Voronoi cells, which delimit the area of influence of each transceiver tower or antenna. 
The Voronoi tessellation partitions the plane into polygonal regions, 
associating each region with one transceiver tower. The partition is such that all points within a 
given Voronoi cell are closer to its corresponding tower than to any other tower in the map.   

Figure 1 shows activity maps for aggregated data corresponding to a 1-hour interval. The upper panel 
shows the activity pattern (in log$_{10}$ scale) for a peak hour (Monday noon), 
while the lower panel shows the same urban neighborhood 
during an off-peak hour (Sunday at 9 am). The differences between both panels reflect the intrinsic rhythm 
and pulse of the city: we can expect call patterns during peak hours to be dominated by the hectic activity 
around business and office areas, whereas other, presumably residential and leisure areas can 
show increased activity during off-peak times, thus leading to different, spatially distinct activity patterns. 
Besides different spatial patterns, each particular time of the day, as well as each day of the week, 
is characterized by a different overall level of activity. This phenomenon is shown by the plot at the center 
of Figure 1, in which aggregated data for a country is shown as a function of time (data was binned 
in time intervals of 1 hour). As expected, the overall normalization of the aggregated pattern is lower 
during weekends than during weekdays, except around weekend midnights and early mornings, when many people go out.

\begin{figure}[t!]
\includegraphics[width=5.8truein, height=4.2truein]{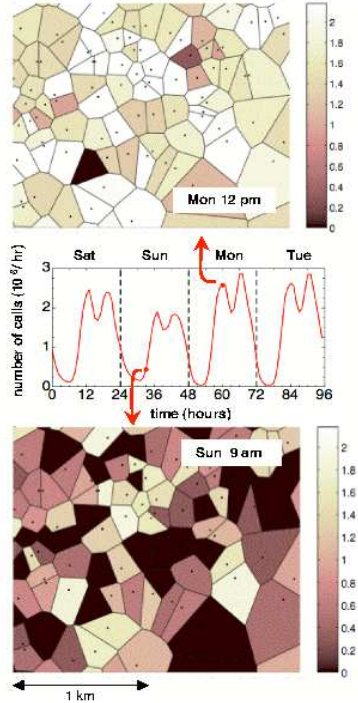}
\caption{Call activity maps in an urban neighborhood, showing the number of calls per hour 
managed by each transceiver tower or antenna (dots). The division in terms of Voronoi cells defines 
the area of reach of each tower. Call traffic patterns depend on time and day of the week, as shown by 
comparing the map on a Monday at noon (upper panel) with that on a Sunday at 9 am (lower panel). The bars 
on the right side of each panel correspond to the number of calls per hour and tower in log$_{10}$ scale.}
\label{fig1}
\end{figure}

The minimum spatial resolution is determined by either the typical distance between towers or, in  
rural regions with sparse tower density, by the reach of the radio-frequency signals exchanged between the 
mobile handset and the antenna (typically ranging from a few hundred meters to several kilometers). 
To explore activity differences at larger scales, the data of neighboring cells can be aggregated. 
At the expense of some loss of spatial resolution, aggregating data into larger spatial bins (taking, e.g., 
a regular spatial grid covering the entire country) allows for better statistics and for a more stable 
activity pattern. That is, the number of calls made from a group of nearby cells at a certain time 
and day of the week is expected to be fairly constant, except for small statistical fluctuations. 

Usually, activity patterns are strongly correlated with the daily pulse of populated areas (such 
as those shown in Fig. 1) and, at a larger scale, to variations in population density between different 
regions within the country. In contrast, departures from the mean expected activity are in general not 
trivially correlated with population density and describe instead interesting dynamical features. 

The measurement of fluctuations around the mean expected activity is of paramount importance, since 
it allows a quantitative measurement of anomalous behavior and, ultimately, of possible emergency  
situations. This indeed constitutes the base of proposed real-time monitoring tools such as the 
{\it Wireless Phone-based Emergency Response} (WIPER) system \cite{mad06}. Anomalous patterns indicative of 
a crisis (such as the occurrence of natural catastrophes and terrorist attacks) could be detected 
in real time, plotted on satellite and GIS-based maps of the area, and used in the immediate evaluation of 
mitigation strategies, such as potential evacuation routes or barricade placement, by means of computer 
simulations \cite{mad06,sch07}. 

\begin{figure*}[t]
\begin{center}
\includegraphics*[width=10.0cm]{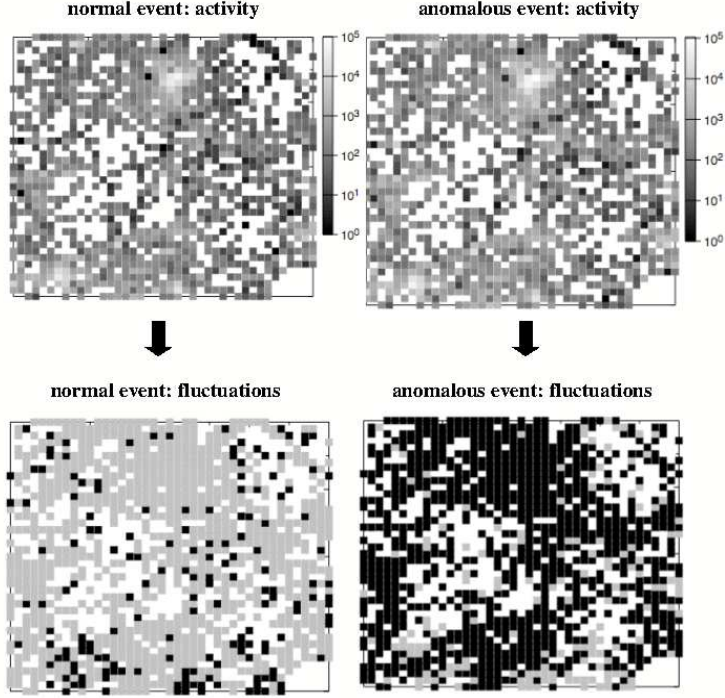}
\end{center}
\caption{Activity and fluctuations in a regular 2D grid showing a normal event (left panels) and an 
anomalous one (right panels). The activity is displayed in terms of the number of calls 
per hour inside each square bin in log$_{10}$ scale (upper panels). High-activity bins above the 
fluctuation threshold $A_{thr}=0.25$ are shown in black, while bins with normal activity are 
shown in grey (bottom panels). Bins in white correspond to areas not covered by the 
mobile phone carrier.}
\label{fig2}
\end{figure*}
       
The call volume shows strong variations with time and day of the week, as shown in Figure 1, but 
differences across subsequent weeks are generally mild (provided one considers call traffic in the 
same place, time and day of the week). 
To capture the weekly periodicity of the observed patterns, 
we define $n_i({\bf{r}},t,T)$ as the number of calls recorded 
at location ${\bf{r}}$ (which can either denote a single Voronoi cell or a group of neighboring cells) 
during the $i$th week between times $t$ and $t+T$, where time is 
defined modulo 1 week. 
Assuming we have access to continuous data for $N$ weeks, the mean call activity is given by
\begin{equation}
\langle n({\bf{r}},t,T)\rangle = {{1}\over{N}}\sum_{i=1}^Nn_i({\bf{r}},t,T)\ .
\end{equation} 
Note that, in the same way as one can trade off spatial resolution for increased 
statistics by summing over a group of Voronoi cells, varying $T$ one can regulate time accuracy versus 
statistics. This certainly depends on the extent to which aggregated data shows a regular, stable 
behavior. The results presented here correspond to $T=1$ hour. 

\begin{figure}[t]
\centerline{{\epsfxsize=5.2in \epsfysize=2.3in \epsfbox{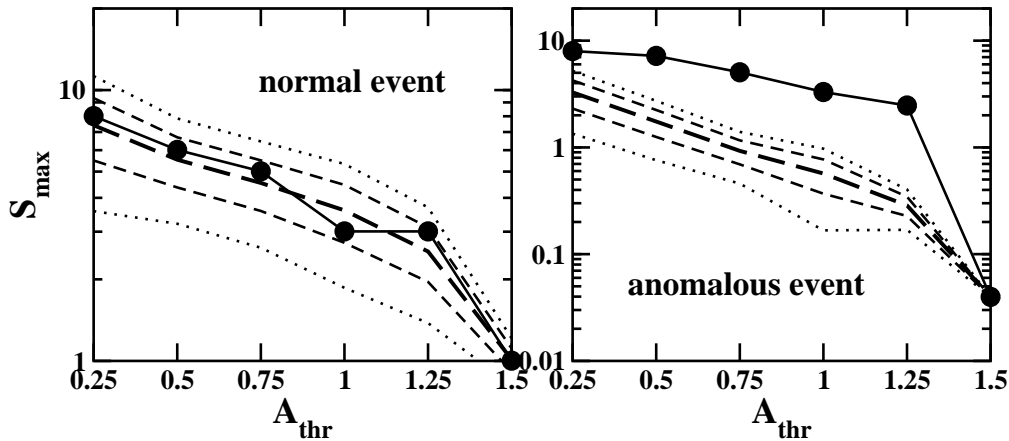}}}
\caption{Size of the largest cluster as a function of the fluctuation 
threshold for the normal case (left) and the anomalous one (right). Measurements on the call data 
(solid line with circles) are compared to those of randomized distributions, of which we show  
the mean (long-dashed line) and confidence bounds at $\pm\sigma_{rdm}$ (short-dashed lines) 
and $\pm 2\sigma_{rdm}$ (dotted lines).}
\label{fig3}
\end{figure}

The scale to measure  
departures from the average behavior is set by the {\it standard deviation}, defined as 
\begin{equation}
\sigma({\bf{r}},t,T) = \sqrt{{{1}\over{N-1}}\sum_{i=1}^N
\left(n_i({\bf{r}},t,T)-\langle n({\bf{r}},t,T)\rangle\right)^2}\ .
\end{equation}   
Hence, using recorded data for an extended period of time, one can determine the expected call traffic 
levels and corresponding deviations for all times and locations. Once this {\it normal} behavior is established, 
{\it anomalous} fluctuations above or below a given threshold can be obtained using the condition
\begin{equation}
|n_i({\bf{r}},t,T)-\langle n({\bf{r}},t,T)\rangle| > A_{thr}\times\sigma({\bf{r}},t,T)\ ,
\end{equation}      
where $A_{thr}>0$ is a constant that sets the fluctuation level.    

We grouped Voronoi cells together generating a regular 2D grid made of square bins of 
about 12 km of linear size. Considering a fixed time slice, we study 
the spatial clustering of bins showing anomalous activity at different fluctuation levels.
In order to illustrate our procedure, Figure 2 shows the activity and fluctuations in a grid of 
size $40\times 40$ bins (i.e. $480\times 480$ km$^2$ area). 
We compare the activity in the same region for 2 different weeks (corresponding to the 
same time and day of the week). The left panels show a {\it normal event}, in which fluctuations 
around the local mean activity are typically small, with just a few scattered bins having somewhat larger 
deviations. The right panels, however, show an {\it anomalous event}, characterized by extended, spatially 
correlated fluctuations that indicate the emergence of a large-scale, coordinated activity pattern. As pointed 
out above, the existence of anomalous activity patterns could be indicative of possible emergency situations.  
Similarly to the Voronoi maps already discussed, the upper panels in Fig.2 show the activity (number of calls 
per hour inside each square bin) in log$_{10}$ scale. White bins correspond to areas not covered by the 
mobile phone provider. Taking a fixed threshold value $A_{thr}=0.25$, 
the bottom panels show the high-activity bins above the fluctuation threshold (in black) and the bins with 
normal activity (in grey). Note that, although the activity maps have a similar appearance to the degree 
that they seem at first look indistinguishable, the fluctuation maps 
display striking differences.  

\begin{figure}[t]
\centerline{{\epsfxsize=5.2in \epsfysize=2.3in \epsfbox{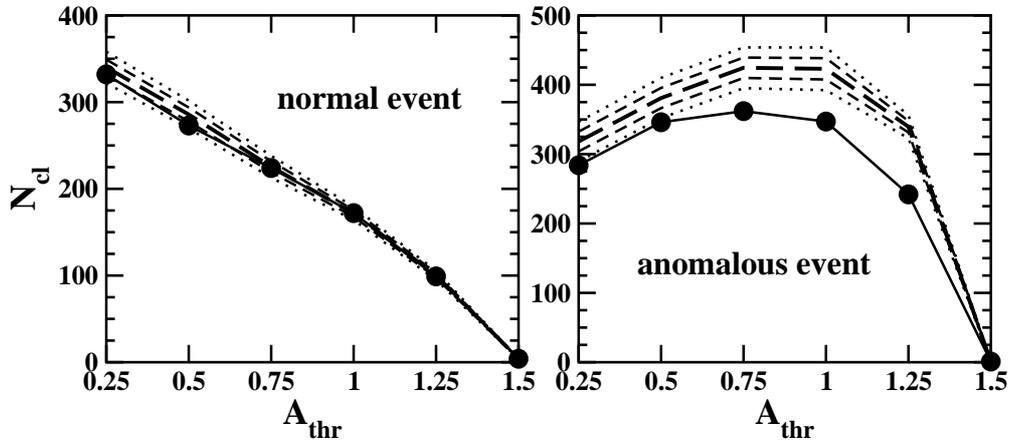}}}
\caption{Number of different clusters as a function of the fluctuation 
threshold for the normal case (left) and the anomalous one (right). 
Measurements on the call data (solid line with circles) 
are compared to results on random configurations (dashed and dotted lines).}
\label{fig4}
\end{figure}

In order to quantify the clustering of anomalous bins, we will use the standard tools of percolation theory and 
determine the size of the largest cluster, the number of 
different clusters, and the size distribution of all clusters. 
The statistical significance of the measured clustering is evaluated by comparing it to results from 
randomized distributions, in which 
many different configurations are randomly generated, keeping fixed the total number of high-activity  
bins above the fluctuation threshold. The substrate, which is formed by all bins with non-zero activity, 
remains always the same (in Fig.2, for instance, the substrate is the set of all grey and black bins). Clusters 
are defined by first- and second-order nearest neighbors in the square 2D grid.    
In the remainder of this section, we will focus on a specific large-scale anomalous event and 
compare it to the normal behavior 
observed in data of a different week (but corresponding to the same time and day of the week). The comparison 
between normal and anomalous events will illustrate the use of percolation observables as diagnostic tools 
for anomaly detection. 

\begin{figure}[t]
\centerline{{\epsfxsize=5.2in \epsfysize=3.2in \epsfbox{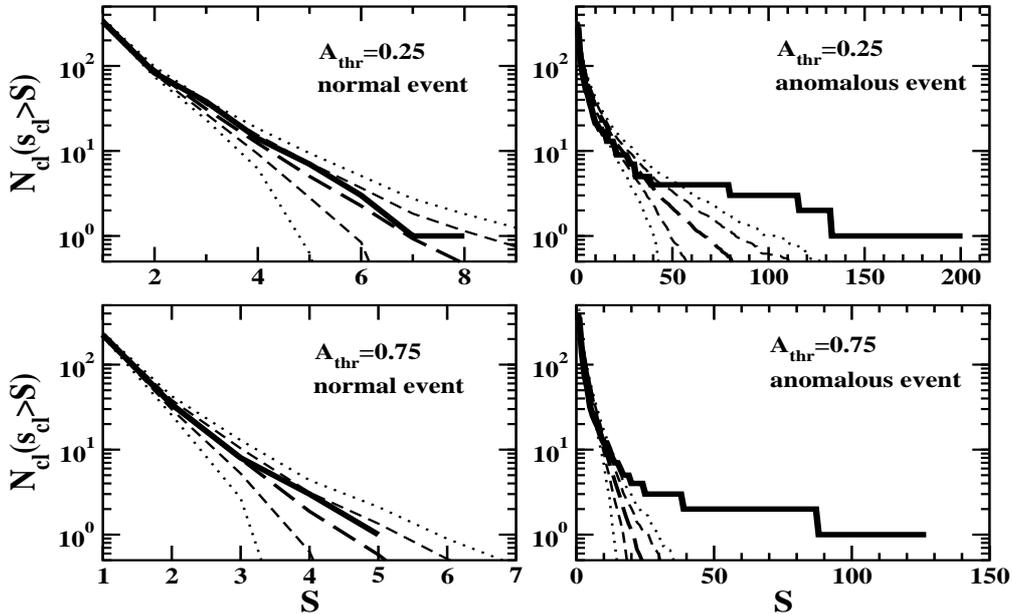}}}
\caption{Cumulative size distribution of all clusters as a function of cluster 
size, for $A_{thr}=0.25$ (upper panels), $A_{thr}=0.75$ (bottom panels), normal case 
(left panels), and anomalous case (right panels). Thick solid lines are measurements on the call data, 
while dashed and dotted lines are results from random configurations.}
\label{fig5}
\end{figure}

Figure 3 shows the size of the largest cluster, $S_{max}$, as a function of the fluctuation 
threshold $A_{thr}$, for the 
normal case (left) and the anomalous one (right). 
Each measured plot (solid line with circles) is compared to results from randomized 
distributions. The latter correspond to the mean (long-dashed line) 
and confidence bounds at $\pm\sigma_{rdm}$ (short-dashed lines) 
and $\pm 2\sigma_{rdm}$ (dotted lines), as obtained from generating 100 random configurations in each case. 
As expected, the plots show that the size of the largest cluster monotonically decreases with the fluctuation threshold. 
However, while the clustering in the normal case lacks any significance, the anomalous event shows large departures 
from the clustering expected in a random configuration.       

In the same vein, Figure 4 shows the number of different clusters, $N_{cl}$, as a 
function of the fluctuation threshold $A_{thr}$, where measurements on the call data for the same
normal (left) and anomalous (right) events are compared to results from randomized configurations. 
As before, in the normal case the number of clusters agrees well with the expectations for random configurations, 
while significant departures are observed in the anomalous case. 

Figure 5 shows the cumulative size distribution of all clusters, $N_{cl}(s_{cl}>S)$, as a function of the cluster 
size $S$, compared to random configurations. The upper panels display results for $A_{thr}=0.25$, while 
the bottom ones show results for $A_{thr}=0.75$, as indicated. 
Moreover, the left panels correspond to the normal event, 
while the right panels to the anomalous event. Again, the measured cluster size distribution in the normal 
case is in good agreement with the expected one for a random configuration. In contrast, the anomalous event shows the 
occurrence of a few very large clusters formed by many highly active bins. 
These unusually large structures cannot be explained as arising just from random configurations, but instead are 
the result of the spatiotemporal correlation of large, highly active regions. 

As a summary, in this Section we showed how large-scale collective behavior 
can be described using aggregated data resolved in both time and space. 
Moreover, we developed the basic framework 
for detecting and characterizing spatiotemporal fluctuation patterns, 
which is based on standard procedures of statistics and percolation theory.  
These tools are particularly effective in detecting extended anomalous events, 
as those expected to occur in emergency scenarios due to e.g. natural 
catastrophes and terrorist attacks. 

\section{Individual calling activity patterns}
In order to use the huge amount of data recorded by
mobile phone carriers to investigate 
various aspects of human
dynamics~\cite{gon07,laszlo,list41,list42,list43},
a necessary starting point it is to characterize
the dynamics of the individual calling activity {\it per se}.  
Previous studies have measured the time between consecutive
individual-driven events, such as sending e-mails, printing,
and visiting web pages or the library~\cite{Olivera,Maya}. 
Those events are described by heavy-tailed 
processes~\cite{laszlo,Goh}, challenging the traditional Poissonian modeling 
framework~\cite{VazquezPRL1,Caldarelli,Blanchard,Daly,Cesifoti}, with consequences on
task completion in computer systems.
In this section we explore the interevent distribution
of the calling activity of $6 \times 10^{6}$ mobile phone users during $1$
month.

\begin{figure*}[t]
\begin{center} 
\includegraphics*[width=10.0cm]{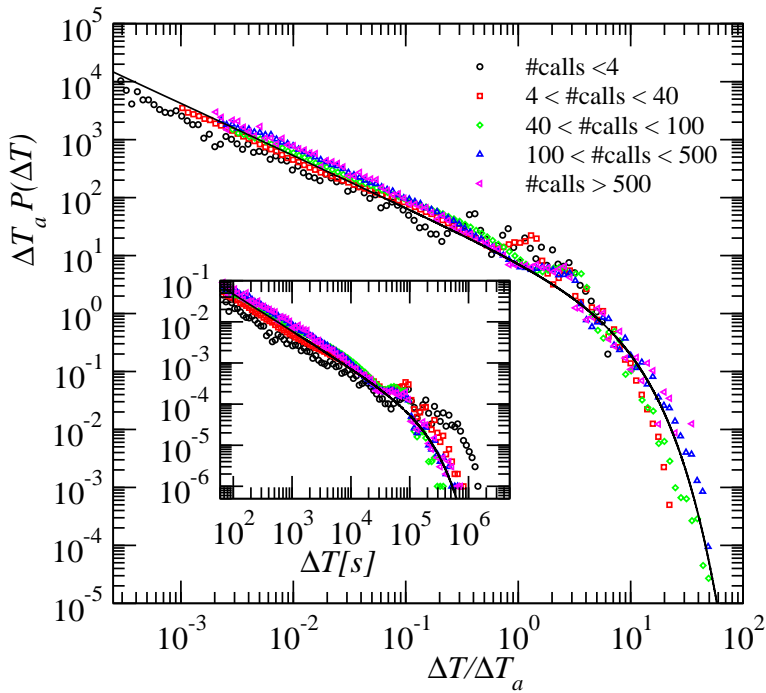}
\end{center}
\caption{\protect Interevent time distribution 
$P(\Delta T)$ for calling activity. 
$\Delta T$ corresponds to the time interval
between two mobile phone calls sent by the same user.
Different symbols indicate the measurements done over
groups of users with different activity levels (\# calls).
The inset shows the unscaled interevent time
distribution and the solid line corresponds to
Eq.~(\ref{eq:distr}).}  
\label{fig6}
\end{figure*}
 
As many other human activities, the calling activity 
pattern is highly heterogeneous.
While some users rarely use the mobile phone, others
make hundreds or even thousands of calls each month.
To analyze such different levels of activity, we group
the users based on their total number of calls.  
Within each group, we measure the probability density 
function $P(\Delta T)$ of the time interval  $\Delta T$ between 
two consecutive calls made by each user. 
As shown by the inset of Fig.~\ref{fig6}, the tail of the distribution
is shifted to longer interevent times for users with less activity.
However, if we plot $\Delta T_{a} P(\Delta T)$ as a function 
of $\Delta T/\Delta T_{a}$, where $\Delta T_{a}$ is
the average interevent time for the corresponding user, the data collapses into a 
single curve (Fig.~\ref{fig6}). This
indicates that the measured interevent distribution follows the expression 
$P(\Delta T)$ = $1/\Delta T_{a}$$\mathcal{F}$$(\Delta T / \Delta T_{a})$, where 
$\mathcal{F}$$(x)$ is independent from the average
activity level of the population. This represents a
universal characteristic of the system that surprinsingly also
coincides with results from e-mail communication~\cite{cond-matGoh}.
The data are well fitted by
\begin{equation}
P(\Delta T) = (\Delta T)^{-\alpha} \exp (\Delta T/\tau_{c}), 
\label{eq:distr}
\end{equation}
where the power law scaling with exponent $\alpha = 0.9 \pm 0.1$ is followed
by an exponential cutoff at $\tau_{c} \approx 48$ days. 
Equation (\ref{eq:distr}) is shown by a solid line in the inset
of Fig.~\ref{fig6} and its scaled version is presented
in the main panel of the figure using $\Delta T_{a}=8.2$ hours,
which is the average interevent time measured for the whole population. 
This result, clearly different from the one predicted
by a Poisson approximation~\cite{Goh,Feller,Sornette}, would for
instance affect the predictions of spreading dynamics 
through the network of calls~\cite{VazquezPRL2}.

\begin{figure*}[t]
\begin{center}
\includegraphics*[width=10.0cm]{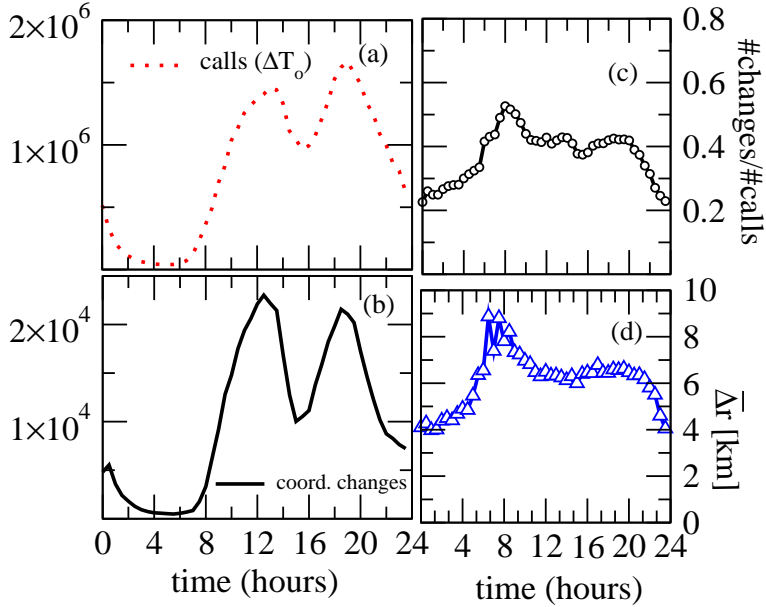}
\end{center}
\caption{\protect Travel behavior. {\bf (a)-(b)} Number of trips 
and consecutive calls that are reported within a fixed interevent
time $\Delta T_{o}=30$ min vs. time of the day.
{\bf (c)} The ratio of the two quantities described 
in (a) and (b) shows that along the whole day $40 \pm 20 \%$ 
of the people that is calling seems to be also traveling.
{\bf (d)} The average distance of travel within $\Delta T_{o}=30$ min
remains constant during the day within $6 \pm 2$ km,
a reasonable value that may correspond to the combination between 
walk and motor transportation.}  
\label{fig7}
\end{figure*}

To explore the interplay between human activity and
mobility patterns, we fix the characteristic observation
time to $\Delta T_{o} = 30$ min and collect only those 
consecutive calls that occur with this
interevent time, recording also the time of the day 
in which they occurred (Fig. \ref{fig7} a). 
For each pair of calls, we count how many of them 
result in a change of coordinate, e.g. the user
traveled in the $30$ min time interval between the calls 
(Fig. \ref{fig7} b). The number of events that result in a change
of location and the number of calls as a function of time capture the daily 
activity pattern of the users~\cite{Huberman}.
We find that both the call and the mobility pattern
decrease at night and have clear peaks near noon and late
evening. There is a factor of $30$ between the largest and the smallest
number of events (calls/changes of location) 
reported during the day. Interestingly, when  
we calculate the fraction of consecutive calls also resulting
in a potential change of location, the quantity varies 
at most $40 \%$ during the whole
day (Fig.~\ref{fig7}c). This indicates that although the total activity
varies strongly, the percentage of the people
that are calling and traveling remains rather stable.
More importantly, the average distance traveled within $\Delta T_{o} = 30$
min. is stable in the vicinity of $\Delta r = 6 \pm 2$ km 
(Fig.~\ref{fig7}d), a value consistent for the combination between 
walk and motor transportation.

\section{Conclusions}

Novel aspects of human dynamics and social interactions were addressed by means of 
mobile phone data with time and space resolution. 
This allowed us to study the mean collective behavior at 
large scales and focus on the occurrence of anomalous events. 
Considering a fixed time slice,    
we partitioned the space using a regular grid and studied the aggregated call activity inside each 
square bin forming the grid.
We showed that anomalous events give rise to spatially extended patterns that can 
be meaningfully quantified in terms 
of standard percolation observables. 
By considering a series of consecutive time slices, we could investigate the 
rise, clustering and decay of spatially extended anomalous events, which could be 
relevant e.g. in real-time detection of emergency situations. 
  
We also investigated patterns of calling activity at the individual level. 
We observed that 
the interevent time of consecutive calls is heavy-tailed,  
a finding that has implications for dynamics of spreading phenomena on social networks, and that 
agrees with results previously reported on other, related human activities. 
We also show that, despite of the complexity inherent in the
interevent calling patterns, 
it is still possible to recover some 
characteristic values from the behavior 
of the population that are stationary during the day, 
such as the fraction of active traveling population 
and their average distance traveled.

In many ways, these results represent only a first step towards understanding human activity patterns. 
Our results indicate that the rich information provided by mobile communication data
open avenues to addressing novel problems. These tools offer a chance to improve our 
understanding of complex networks as well \cite{CNP1,CNP2,CNP3,CNP4,CNP5,CNP6,CNP7,CNP8}, 
by potentially correlating the structure of social 
networks with the spatial layout of the users as nodes \cite{SNP1,SNP2,SNP3,SNP4,gon_prl,lind1,lind2}, 
thus contributing to a better understanding 
of the spatiotemporal features of network evolution.

\section*{Acknowledgments}
This work was supported by the James S. McDonnell Foundation 21st Century Initiative in 
Studying Complex Systems, the NSF within the DDDAS (CNS-0540348), 
ITR (DMR-0426737) and IIS-0513650 programs, as well as by U.S. Office of Naval Research N00014-07-C 
and the NAP Project sponsored by the National Office for Research and Technology (KCKHA005). 
Data analysis was performed on the Notre Dame Biocomplexity Cluster
supported in part by NSF MRI Grant No. DBI-0420980.

\end{document}